\documentclass{optica-article}

\journal{opticajournal} % for journals or Optica Open

\articletype{Research Article}

\usepackage{lineno}
\usepackage{amsmath}
\DeclareMathOperator*{\argmin}{argmin}
\newcommand{\uled}[0]{$\mu$LED}
\newcommand{\uleds}[0]{$\mu$LEDs}
\newcommand{\figref}[1]{Fig.~\ref{#1}}
\newcommand{\pfigref}[1]{Figure~\ref{#1}}
\newcommand{\secref}[1]{Sec.~\ref{#1}}

\begin{document}

\title{Fourier modal method for inverse design of metasurface-enhanced micro-LEDs}

\author{Martin F Schubert\authormark{1,*} and Alec M Hammond\authormark{1}}

\address{\authormark{1}Meta, 1 Hacker Way, Menlo Park, CA 94025, USA}

\email{\authormark{*}mfschubert@gmail.com} 

\begin{abstract*} 
We present a simulation capability for micro-scale light-emitting diodes (\uled{}s) that achieves comparable accuracy to CPU-based finite-difference time-domain simulation but is more than $10^7$ times faster. Our approach is based on the Fourier modal method (FMM)---which, as we demonstrate, is well suited to modeling thousands of incoherent sources---with extensions that allow rapid convergence for \uled{} structures that are challenging to model with standard approaches. The speed of our method makes the inverse design of \uled{}s tractable, which we demonstrate by designing a metasurface-enhanced \uled{} that doubles the light extraction efficiency of an unoptimized device.
\end{abstract*}

%%%%%%%%%%%%%%%%%%%%%%%%%%  body  %%%%%%%%%%%%%%%%%%%%%%%%%%

\section{Introduction}

Micro-scale light-emitting diodes (\uled{}s) with lateral dimensions near 1 $\mu$m are of interest for a variety of applications including augmented reality displays \cite{Xiong2021, Park2021}. In this context, high light extraction efficiency (LEE) is essential due to the requirement for high brightness (e.g. for outdoor usage) and the need to operate for extended periods on battery power. It would be desirable to apply methods such as inverse design to obtain \uled{} designs with higher performance; inverse design is a powerful technique that automatically discovers the topology and shapes of designs that minimize some objective function \cite{Molesky2018, meep_adjoint, hammond_constraints}, and would be suitable to the creation of metasurfaces that enhance \uled{} LEE. However, light generated in the \uled{} is spatially incoherent; spatially incoherent sources (arising in e.g. spontaneous emission \cite{MILONNI19761, KIM1986396, POLIMERIDIS2015} and thermal emission \cite{Carey2008, Basu2009}) are extremely expensive to model by standard approaches (requiring many independent simulations), making inverse design of structures including incoherent sources computationally intractable. Consequently, previous works on modeling and optimization of \uled{}s have focused on symmetric or low-dimensional cases, or used a widely-spaced dipoles to approximate a planar active region \cite{cylindrical_uled_fdtd, Chung:22}, leaving the challenge of general 3D \uled{} inverse design unaddressed.

In this work, we present a new simulation capability based on the Fourier modal method (FMM) which overcomes these limitations. For the \uled{}, our method gives results that are in excellent agreement with finite-difference time-domain (FDTD) simulations while being more than $10^7$ times faster~(\secref{sec:validation}). This makes it practical for use in an inverse design setting, which we demonstrate by designing metasurface-enhanced \uled{}s that significantly improve LEE~(\secref{sec:inverse_design}). The speed of our method also enables calculation of quantities such as high-resolution spatial maps of LEE, allowing new insights into the physics of \uled{} devices.

Related to our work, novel factorization methods for multi-channel inverse design~\cite{multisource, li2023fast} and a trace formulation of photonic inverse design~\cite{Yao2022} have recently been introduced, which have lower computational cost compared to conventional inverse design formulations. Our work shows that---for problems such as the \uled{}---with an appropriate choice of the simulation algorithm, the standard formulation may be sufficient. In particular, the FMM has several characteristics that enable efficient modeling of many incoherent sources. The FMM treats fields in periodic stratified media in a truncated Fourier basis, which has the potential to represent fields accurately with relatively few terms. As in other methods, with the FMM one solves a linear system to obtain the fields generated by a source; this system has dimension given by the size of the basis, which for FMM is related only to the details of the structure in \emph{two} dimensions, as the third dimension is handled analytically. The resulting linear system can be relatively small, enabling direct solution methods to be employed. Finally, while constructing the linear system involves eigendecompositions and is expensive, the results of these operations can often be reused e.g. in an optimization setting where many layer profiles may be unchanged between iterations.

Despite these potential advantages, there are challenges that have discouraged the use of the FMM for problems such as the \uled{}. For example, the original and common FMM formulation exhibits poor convergence in structures containing metals \cite{Li:93}. In addition, FMM naturally treats periodic structures and sources; while \uled{}s are generally formed in periodic arrays, treatment of a \emph{source} as periodic gives rise to nonphysical interference effects \cite{BZ_integration_1}. We address these issues with a vector formulation of the FMM---including an improved method for automatically computing vector fields---and Brillouin zone integration \cite{BZ_integration_1, BZ_integration_2, BZ_integration_3}, which dramatically improve convergence (\secref{sec:convergence}) and allow modeling of localized dipoles in a periodic \uled{} array.

Our implementation of the FMM is {\tt FMMAX}, which is based on \emph{Jax} \cite{jax2018github} and joins several recent codes that support automatic differentiation \cite{KIM2023108552, Colburn:21, Jin2020, nannos}, enabling the gradient calculation needed for inverse design. Our code is distinguished by a flexible low-level programming interface which admits uncommon use cases such as the \uled{} simulations carried out here, and is freely available at \href{https://github.com/facebookresearch/fmmax}{github.com/facebookresearch/fmmax}.

\section{Method}
\label{section:method}

Throughout this section, we discuss extensions of the basic FMM needed to enable \uled{} simulation and inverse design. For a full description of the FMM, we refer the reader to \cite{LIU20122233} and \cite{Whittaker:09}, which our implementation closely follows. 

\subsection{Vector FMM formulations} \label{sec:vector_form}
It has long been recognized that the original FMM formulation (referred to as \emph{FFT}, as in \cite{LIU20122233}) converges well only when electric fields are tangent to material interfaces \cite{Li:93}; this was improved by vector FMM formulations, which introduce a local coordinate system with unit vectors that are tangent and normal to the boundaries of features. Vector formulations allow orientation-dependent Fourier factorizations of permittivity to be used for tangent and normal components of the electric displacement field, which is the essential change of the vector FMM methods from the original formulation \cite{Li:96}. In an inverse design setting, the vector fields must be automatically generated from the evolving geometry. Various schemes have been developed based on interpolation \cite{Gotz:08} and functional minimization \cite{LIU20122233}; we adopt the latter due to its applicability to continuously varying permittivity, as is encountered during the course of optimization. 

%Generation and normalization of vector fields has been the subject of several works \cite{nannos, LIU20122233, Granet:96, Li:96, Li:97, David:06, Schuster:07, Gotz:08, Antos:09}.

As in \cite{LIU20122233}, the vector field $\mathbf{t} = \left[t_x, t_y\right]^\intercal$ for a permittivity array $\varepsilon$ is obtained by minimizing a loss function $\mathcal{L}$. High-quality vector fields should be smoothly varying and normal to the permittivity gradient, i.e. tangent to the interfaces in the structure. 
This is achieved using
\begin{align}
\label{eqn:vector_loss}
\mathcal{L}(\mathbf{t}, \varepsilon) = \sum \left|t_x \nabla_y \varepsilon - t_y \nabla_x \varepsilon \right| + \left|\nabla_{xy} t_x\right|^2 + \left|\nabla_{xy} t_y\right|^2,
\end{align}
where the sum is over all the elements in $\mathbf{t}$. 
The first term is minimized when $\mathbf{t}$ is correctly oriented, while the remaining terms are minimized when $\mathbf{t}$ varies slowly. The resulting optimization problem,
\begin{align}
\label{eqn:vector_optimization}
\mathbf{t} = \argmin_{
\substack{\mathbf{t}^\star \\ \text{s.t.} \ \left|\mathbf{t}^\star\right| \leq 1}}
\mathcal{L}(\mathbf{t}^\star, \varepsilon),
\end{align}
can be solved by standard gradient-based methods.

The choice of initial $\mathbf{t}^\star$ is consequential. In a novel method termed \emph{Jones direct}, we compute an initial real-valued vector field (even for complex $\varepsilon$) and convert to a complex Jones field using the method of \cite{Antos:09}. The final complex-valued $\mathbf{t}$ is obtained by directly optimizing this Jones field, and advantageously lacks both discontinuities and zeros.

A real-valued $\mathbf{t}$ can be obtained by avoiding the Jones conversion and directly optimizing a real $\mathbf{t}^{\star}$; this will lack discontinuities but include zeros, and is equivalent to the \emph{Pol} method of \cite{LIU20122233}. The field can be normalized to have magnitude 1 everywhere (the \emph{Normal} method of \cite{LIU20122233}) or it can be converted to a Jones field as a post-processing step (the \emph{Jones} method of \cite{LIU20122233}). In general, the \emph{Normal} method will yield fields having discontinuities, while the \emph{Jones} method will yield fields that lack discontinuities and zeros, but vary more rapidly than those obtained by the \emph{Jones direct} method. Thus, the \emph{Jones direct} scheme has favorable characteristics for generation of a local coordinate system represented in a Fourier basis.

Example vector fields for these methods are in Appendix A. Consistent with the theoretical benefits discussed above, we observe best convergence for the FMM with the \emph{Jones direct} formulation (Appendix B) and use the method throughout this work.

% The \emph{FFT} method of \cite{LIU20122233}---which is the standard formulation supported by the majority of FMM codes \cite{KIM2023108552, Jin2020, Colburn:21}---is also implemented.

\subsection{Brillouin zone integration for aperiodic sources} \label{sec:bz_integration}
Sources in periodic structures can be modeled in the FMM by expanding the source spatial profile in terms of the layer eigenmodes \cite{Whittaker:09}. However, in this approach both the sources and the structure are periodic, which is not the case in a real $\mu$LED and can give rise to unphysical interference effects.

In principle, larger supercells including multiple $\mu$LEDs could be modeled, but due to the unfavorable scaling of the FMM algorithm this may be prohibitive. Absorbing boundary conditions could also be introduced, as used previously in extensions of the FMM to accommodate aperiodic structures \cite{Lalanne:00, Silberstein:01}. A further alternative---the approach adopted here---is the Brillouin zone (BZ) integration method of \cite{BZ_integration_1, BZ_integration_2, BZ_integration_3}, in which the electric field from an isolated source in a periodic structure is found by,
\begin{align}
E(r) = \frac{1}{A_{BZ}} \int_{BZ} E_{k_B}(r) d k_B,
\end{align}
where $E_{k_B}$ is the field with Bloch-periodic boundary conditions for wavevector $k_B$, and $A_{BZ}$ is the BZ area. The calculation of each $E_{k_B}$ requires a simulation of a single unit cell, and thus the problem of modeling a large supercell is transformed into one of several smaller simulations.

Throughout this work, we approximate integrals over the Brillouin zone by averaging over a regular $k_B$ grid. When the grid has shape $M\times M$, we refer to the case as $M\times M$ BZ integration; when no BZ integration is performed, we refer to this as the periodic dipole approximation (PDA). 

\subsection{Calculation of light extraction efficiency}
For the $\mu$LED modeled here, we are primarily concerned with the total LEE and the power radiated by the dipole. For a given wavelength, the LEE is obtained by,
\begin{align}
\label{eqn:lee}
LEE = \sum_{r, p, k_B} w_p P_{r, p, k_B}^{extracted}\bigg/\sum_{r, p, k_B} w_p P_{r, p, k_B}^{emitted}
\end{align}
Here, $P^{emitted}_{r, p, k_B}$ and $P^{extracted}_{r, p, k_B}$ are the power emitted by a dipole, and power emitted by the dipole which is extracted from the \uled{}. The powers are measured with planar monitors, obtained by summing over the powers in all the Fourier orders as described in \cite{LIU20122233}, and are indexed by $r$, $p$, and $k_B$, corresponding to the dipole position, dipole orientation, and BZ point, respectively. We use a dipole-orientation-dependent weight $w_p$, with values of 1 for the $x$- and $y$-oriented dipoles and 0.1 for the $z$-oriented dipoles. These generally have lower extraction efficiency and can be suppressed in material systems such as AlInGaP by appropriately straining the quantum wells \cite{band_structure_engineering}.

\section{\uled{} structure and optimization problem} \label{sec:optimization}
\begin{figure}[h]
    \centering
    \includegraphics{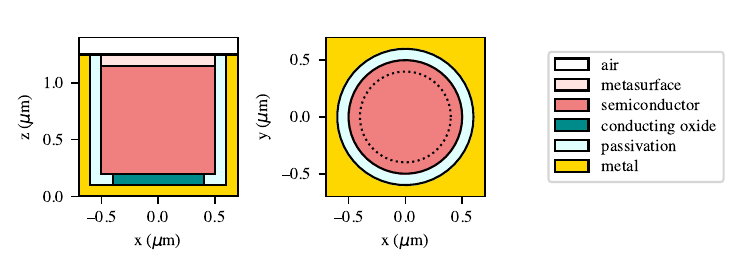}
    \caption{
    \label{fig:structure}
    Side and top view of the $\mu$LED unit cell analyzed throughout this work. A high index semiconductor ($n=3.0$) cylinder is embedded within a metallic substrate ($n=0.2+3.3i$) and separated by a thin passivation layer ($n=1.5$) and transparent conducting oxide ($n=1.9+0.005i$). The metasurface ``design region'' of the \uled{} unit cell resides on top of the semiconductor cavity and emits into air. The material of the metasurface is composed of a refractive index that is linearly interpolated between the index of the semiconductor and air. The metasurface itself is parameterized by a 2D grid of ``pixels''~\cite{sigmund_overview} in $x/y$ and projected into the $z$-direction using a thickness chosen by the optimizer. Dipoles within the \uled{} are located in a planar source region embedded within the semiconductor layer, which is free to move as the optimizer dictates.
    }
\end{figure}

We now describe the \uled{} geometry and corresponding optimization problem used throughout the rest of our analysis. We consider an array of simplified cylindrical \uleds{} with a 1.4 $\mu$m pitch, which is illustrated in \figref{fig:structure}. The semiconductor region has a diameter of 1~$\mu$m, the conducting oxide has a diameter of 0.8~$\mu$m, and the passivation sidewall has a thickness of 0.1~$\mu$m. Other layer thicknesses are optimizable parameters specified elsewhere. A spatial resolution of 10~nm is used for all layers. Refractive indices of the materials are as follows: semiconductor, 3.0; conducting oxide, 1.9 + 0.005$i$; passivation, 1.5; and metal, 0.2 + 3.3$i$. Values are chosen to be representative of AlInGaP, indium-tin oxide, silicon oxide, and gold, respectively.

We model dipoles in a plane within the semiconductor material, and consider $x$-, $y$-, and $z$-oriented dipoles on a 20~nm grid (5847 total dipoles). The spatial profile of the dipoles is Gaussian with a 60 nm full-width at half-maximum. Power emitted by each dipole is measured using planar monitors positioned 50~nm from the dipole plane. We model a single plane, corresponding e.g. to a single quantum well active region, but note that it would be possible to model multi-quantum-well active region with no additional eigendecompositions, and therefore relatively cheaply.

We turn now to the \uled{} optimization problem. In this paper, we exercise unconstrained optimization, as our primary aim is to demonstrate the functioning of the simulator and inverse design pipeline, rather than to obtain designs suitable for manufacture, or to form estimates of the potential performance of metasurface-enhanced $\mu$LEDs. We use a monochromatic objective $\mathcal{L}(p)=1 - LEE(p)$ for a 630~nm emission wavelength. Here, $p$ represents the optimizable parameters, including layer thicknesses and the metasurface density $\rho$. The density is a two-dimensional array with values in the range $[0, 1]$ from which the permittivity at each point is computed by interpolation between the permittivity of air and of semiconductor using the method of \cite{CHRISTIANSEN201923}.

Although one could directly optimize the elements of the density array $\rho$, this can produce designs with very fine features that would not be practical to manufacture. Therefore, we choose to instead optimize a \textit{latent density} $\tilde{\rho}$ also in the range $[0, 1]$, from which $\rho$ is obtained by,
\begin{align}
\rho = \frac{1}{2}\left(1 + \tanh \left(\beta \left(2 \tilde{\rho} - 1\right) \circledast k\right)\right)
\end{align}
where $k$ is a Gaussian kernel with a full-width at half-maximum of 40 nm and $\beta = 2$. This expression is similar to the transforms commonly used in density-based topology optimization \cite{svanberg2013density} and help ensure that $\rho$ lacks features which are too small by encouraging an implicit lengthscale. Futhermore, this approach helps \emph{regularlize} the optimization problem, which is solved using the L-BFGS-B algorithm \cite{lbfgsb}.

\section{Field calculation and validation} \label{sec:validation}
To establish the validity of our method, we first make a direct comparison of the steady-state fields computed by \textit{Meep}~\cite{meep}, a mature FDTD solver, to those computed by FMM. We consider a simple \uled{} lacking a metasurface, with semiconductor and conducting oxide thicknesses of 1.0 $\mu$m and 0.1 $\mu$m, respectively. Dipoles emitting at 630~nm are centered vertically in the semiconductor and positioned (with units of $\mu$m) at $(x, y) = (0, 0)$ and $(0, 0.2)$, both oriented in the $x$-direction.

\begin{figure}
    \centering
    \includegraphics{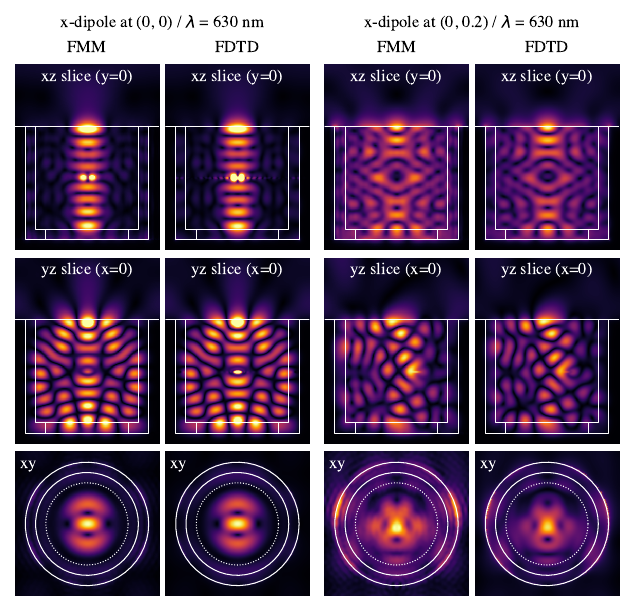}
    \caption{
    \label{fig:field_comparison}
    Cross sections of the steady-state electric field magnitude for dipoles at $(x, y) = (0, 0)$ and $(0, 0.2)$ computed by FMM and FDTD. The $xz$ and $yz$ are taken through the center of the $\mu$LED, while the $xy$ section is taken at the semiconductor/air interface. The methods show strong agreement at all slices and for both dipole positions, as demonstrated by the consistent representation of the complicated modal patterns.}
\end{figure}

\pfigref{fig:field_comparison} shows cross sections of the electric field magnitude computed by the two methods. The field structure for both dipole locations computed by FMM and FDTD are in excellent agreement; in particular, the nodal lines found by the two methods are nearly indistinguishable.

Our FMM calculation uses the PDA with the \emph{Jones direct} formulation and $N=2000$.

The FDTD calculation uses a spatial resolution of 200 voxels / $\mu$m, periodic boundary conditions in the $x$- and $y$-directions, and perfectly electric conductor (PEC) boundary conditions in the z-direction. To mitigate backreflections from the upper z-boundary, we used a 2~$\mu$m thick, adiabatic, conductive absorber layer above the \uled{}. Using a nonlinear optimization algorithm~\cite{nlopt}, we fit a Lorentz-drude material model to the desired complex refractive index values at $\lambda=630$~nm to ensure simulation stability~\cite{taflove}. The FDTD simulation using a broadband Gaussian pulse centered at 630~nm and terminated the simulation after the fields were sufficiently decayed. Just like the FMM solver, the spatial profile of the FDTD source was a normalized Gaussian with a FWHM of 60~nm. The steady-state fields for the 630~nm wavelength were recorded using rolling discrete-time Fourier transform monitors at the specified cross sections.

\section{Solver convergence and performance}\label{sec:convergence}
Next, we study the convergence and performance of the FMM in the context of our \uled{} simulations and compare the results to FDTD. We note that the numerical accuracy of each algorithm is fundamentally dictated by the underlying discretization resolution; for FDTD, this corresponds to the actual spatial resolution of the grid, or the total number of simulation voxels, whereas with FMM this refers to the spatial-frequency resolution, or the total number of Fourier orders. To quantify the relative convergence characteristics of both algorithms, we simulated the \uled{} structure and dipole configuration of \secref{sec:validation} with multiple resolutions.

\begin{figure}[t]
    \centering
    \includegraphics{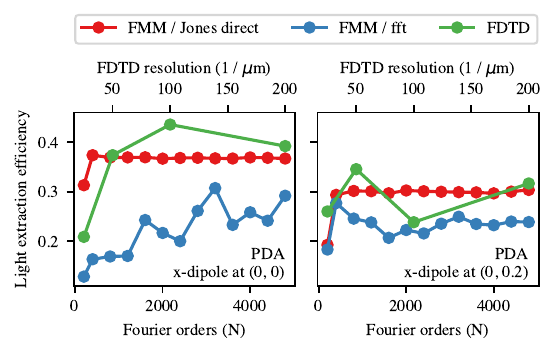}
    \caption{
    \label{fig:convergence_vs_fourier_terms} Convergence of the light extraction efficiency for FMM and FDTD as a function of Fourier order and spatial resolution respectively. Results are given for both the centered and offset $x$-oriented dipole. While the \emph{Jones direct} FMM formulation of FMM converges for $N\geq400$, the \emph{FFT} formulation fails to converge; the calculated efficiency trends in the direction of the \emph{Jones direct} result, but does not reach agreement even for $N=4800$. FDTD requires a resolution of 200 voxels / $\mu$m ($\Delta$x=5~nm) to reach suitable convergence, largely due to the high-contrast metals within the \uled{} model.
    }
\end{figure}

First, we examine the importance of the FMM formulation itself. \figref{fig:convergence_vs_fourier_terms} shows the LEE for the x-oriented dipoles at $(x, y) = (0, 0)$ and $(0, 0.2)$ as a function of the Fourier orders $N$ and for the \emph{Jones direct} and \emph{FFT} formulations. For the \emph{Jones direct} formulation, the LEE for both dipoles quickly converges; the values with $N=4800$ are 36.7\% and 30.4\% respectively, but with only $N=400$ the values are 37.4\% and 29.4\%, in excellent agreement with the converged results. By contrast, the \emph{FFT} formulation exhibits very poor convergence; as $N$ increases the values trend in the direction of the \emph{Jones direct} results, but they exhibit substantial fluctuation and remain far from the converged result even with $N=4800$.

\begin{figure}[t]
    \centering
    \includegraphics{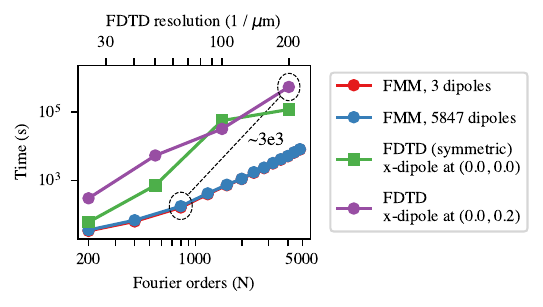}
    \caption{
    \label{fig:speed_test}
    Comparison of elapsed time for FMM and FDTD LEE calculations. The results indicate an approximate $\sim$$3\times10^3$ times speedup when comparing the 200 voxels / $\mu$m FDTD simulation to the $N=800$ Fourier term FMM simulation (resolutions where both algorithms are converged). Notably, the FDTD results correspond to a single dipole simulation. The expected computational cost for additional dipoles scales linearly with the number of dipoles. This is in direct contrast with the FMM method shown here, where the marginal cost of an additional $\sim$6000 dipoles is negligible. Therefore, the projected speedup for a full \uled{} simulation is $\sim$$2\times10^7$.}
\end{figure}

We now directly compare the convergence of FMM and FDTD. \figref{fig:convergence_vs_fourier_terms} also shows the convergence of LEE for both dipole positions calculated by FDTD with varying spatial resolution. With a resolution of 200 voxels / $\mu$m ($\Delta$x=5~nm), we obtain values of 39.2\% and 31.7\%, in good agreement with the converged \emph{Jones direct} result. Such a high spatial resolution for FDTD is expected due to the high-contrast metals within the \uled{} model, although the application of subpixel-smoothing algorithms or other conformal meshing approaches compatible with dispersive media should accelerate the convergence~\cite{deinega2007subpixel}. In combination with the field profiles in \figref{fig:field_comparison}, this shows that the vector FMM can be used for modeling and LEE calculation of \uled{}s, with results comparable to high-resolution FDTD.

Despite the similarity in results between FMM and FDTD, there are major differences in the performance and computational burden of the two algorithms. \figref{fig:speed_test} shows the elapsed time for the FMM and FDTD simulations in \figref{fig:convergence_vs_fourier_terms} as a function of Fourier order and resolution respectively. All calculations were performed using 10 cores of a 2.20GHz Intel$\textsuperscript{\textregistered}$ Xeon$\textsuperscript{\textregistered}$ CPU E5-2698 v4 with 512 GB of RAM and two NUMA nodes. Although {\tt FMMAX} supports GPU acceleration, this was not used to allow a direct comparison. The FMM results are shown for the case where three incoherent dipoles are modeled, and where all $\sim$6000 dipoles are modeled; the results show the negligible incremental cost of simulating additional dipoles. 

The FDTD elapsed times are shown for centered and offset dipoles; the centered-dipole simulation used odd $x$-symmetry and even $y$-symmetry to reduce the computational burden, while the offset dipole considered the full simulation volume and is representative of the general case. Considering this offset dipole, there is approximately a $3\times10^3$ speedup when comparing the 200 voxels / $\mu$m FDTD results (the coarsest resolution that yields converged results) to the $N=800$ Fourier-term results (the lowest $N$ we used in the inverse design setting).

Practically speaking, the speedup is even more significant; if FDTD simulations were carried out for the $\sim$6000 dipoles considered in this work, we project that CPU time would be $\sim$$2\times10^7$ times greater than for the equivalent calculation by FMM. If simulations considering multiple wavelengths were carried out, the advantage would be reduced; FDTD automatically handles multiple wavelengths within a single simulation, while with the FMM each wavelength requires an independent simulation.
However, for the \uled{} we have found that $O(10)$ wavelengths are sufficient, and in any case these simulations can easily be distributed across many machines (the task is embarrassingly parallel). This makes FMM a strong choice for problems such as the \uled{}.

We note a few other important observations regarding the convergence trends of each method. For low $N$, for example, the vector field calculation contributes significantly to the total time of the FMM method, causing the non-polynomial scaling seen in \figref{fig:speed_test}. The FDTD elapsed times also exhibit some deviation from polynomial behavior and a crossover at 100 / $\mu$m resolution. \textit{Meep} requires the user to properly choose a hardware configuration (e.g. the proper number of processes to launch) to maximize the timestepping rate~\cite{oskooi_factorized}. Choosing these parameters correctly is highly dependent on the underlying simulation resolution, and most likely accounts for the aforementioned crossover point. The \uled{} structure may also be more or less resonant for dipoles at different positions in a manner that depends on resolution; for the FDTD algorithm, this can also cause simulation times to vary.

While not clear from the data in \figref{fig:speed_test}, the FMM solver also offers advantages when used within an optimization pipeline where many solves are performed sequentially. Specifically, the eigendecomposition for layers whose cross-sections do not change can be cached and reused. With \emph{Jax}, this is optimization is easily done by \emph{jit}-ing (just-in-time compiling) the simulation and specifying that eigendecompositions be carried out at compile time. For the \uled{} problem, the savings are substantial---while eigendecomposition for three patterned layers is needed at the first step, at subsequent steps only one eigendecomposition (for the metasurface) is needed. For the $N=800$ case on our Xeon$\textsuperscript{\textregistered}$ system, this results in a step time of 49 seconds for the second and subsequent steps, compared to 175 seconds for the first step.

\begin{figure}[t]
    \centering
    \includegraphics{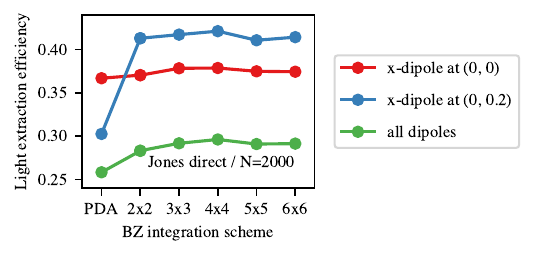}
    \caption{
    \label{fig:convergence_vs_bz_integration}
    Light extraction efficiency for various BZ integration schemes computed with the \emph{Jones direct} FMM formulation and $N=2000$. Values are for individual $x$-oriented dipoles and a full calculation including all $x$-, $y$-, and $z$-oriented dipoles in the basic \uled{}. Each ``scheme'' is defined the regular grid of wavevectors in the first BZ used to approximate a BZ integral, such that $3\times3$ corresponds to nine evenly distributed points (three samples along $k_x$ and three samples along $k_y$). Importantly, for the \uled{} problem, the LEE appears to converge with a relatively small number of points.}
\end{figure}

Finally, to study the effect of BZ integration, we calculated LEE using the PDA and BZ integration schemes up to $6\times6$, using the \emph{Jones direct} formulation and $N=2000$. The results are in \figref{fig:convergence_vs_bz_integration}; values are shown for the two $x$-oriented dipoles and the total LEE considering all $\sim$6000 dipoles in the active region. The quality of the PDA apparently depends on the dipole position: for the dipole at $(0, 0)$, going to $2\times2$ BZ integration and beyond only slightly increases the computed LEE value. By contrast, the dipole at $(0, 0.2)$ has a dramatic increase in the LEE when BZ integration is used, and a sizeable increase is also seen in the all-dipole case. This indicates that avoiding the PDA is critical in the calculation of the LEE for a \uled{}.

Motivated by these results, for the remainder of this work we make use of two simulation configurations. In the inverse design setting, to balance accuracy against speed, we use the \emph{Jones direct} method with $N=800$ and 2$\times$2 BZ integration. For validation, we use $N=2000$, also with \emph{Jones direct} and 2$\times2$ BZ integration.

\section{Optimization results} \label{sec:inverse_design}
To exercise our inverse design pipeline and examine the potential of metasurface-enhanced $\mu$LEDs, we consider three cases: the first is the initial, unoptimized $\mu$LED with structure discussed earlier, i.e. with no metasurface, conducting oxide thickness of 0.1 $\mu$m, semiconductor thickness of 1 $\mu$m, and dipoles are located 0.5 $\mu$m above the oxide. In the second, the film thicknesses are optimized, using the given values as the initial solution. In the third case, both the film thicknesses and the metasurface density are optimizable. Here, the initial semiconductor thickness is reduced to 0.95 $\mu$m and the initial metasurface thickness is 0.1 $\mu$m, resulting in a structure whose optical thickness approximately matches that of the unoptimized \uled{}. The latent density is uniformly initialized with a value of 0.5.

\begin{table}[t]
\centering
\begin{tabular}{ |r|c|c|c| } 
\hline
& unoptimized & thicknesses & thicknesses and metasurface \\ 
\hline
\hline
metasurface & -- & -- & 0.102 $\mu$m \\ 
semiconductor height & 1.000 $\mu$m & 0.945 $\mu$m & 0.966 $\mu$m \\ 
source offset & 0.500 $\mu$m & 0.531 $\mu$m & 0.507 $\mu$m \\ 
conducting oxide & 0.100 $\mu$m & 0.054 $\mu$m &  0.102 $\mu$m \\ 
\hline
\end{tabular}
\caption{Thicknesses for the unoptimized and optimized \uled{}s.}
\label{table:thicknesses}
\end{table}

\begin{figure}[t]
    \centering
    \includegraphics{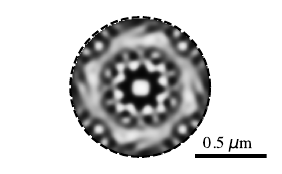}
    \caption{
    \label{fig:metasurface}
    The optimized metasurface density $\rho$; light and dark colors correspond to $\rho=0$ (air) and $\rho=1$ (semiconductor), respectively. The dashed line indicates the passivation/semiconductor interface.}
\end{figure}

The thickness-optimized \uled{} optimization proceeded for 17 iterations before the optimizer converged, reaching an LEE of 40.1\% at 630 nm, compared to 28.4\% for the initial \uled{}. The optimized film thicknesses are given in Table \ref{table:thicknesses}. Changes to thicknesses from the initial values are relatively small, consistent with a loss landscape having many local minima, as frequently encountered in thin film optimization problems. The \uled{} with optimized film thicknesses and metasurface proceeded for 69 iterations, ultimately reaching an LEE of 56.2\%. The film thicknesses are given in Table \ref{table:thicknesses}, and the metasurface density is shown in \figref{fig:metasurface}.

\begin{figure}
    \centering
    \includegraphics{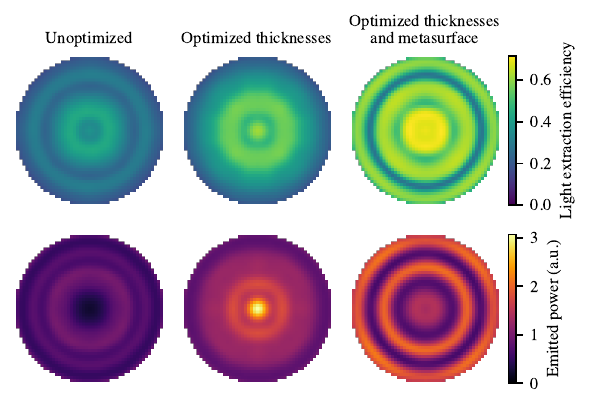}
    \caption{
    \label{fig:lee_maps}
    Spatial maps of light extraction efficiency and emitted power for the initial, thickness-optimized, and metasurface-enhanced $\mu$LED. Values are for 630 nm wavelength, averaged over dipole orientations, and for $N$=2000 with 2$\times$2 BZ integration.}
\end{figure}

To understand the mechanism for LEE improvement in the \uled{}s obtained by optimization, we computed high-resolution maps of the LEE and emitted power for the three \uled{} structures. These are shown in \figref{fig:lee_maps}. As noted, our calculation considers dipoles at $\sim$2000 locations in a plane; each pixel in a map corresponds to a single physical dipole location, with value found by appropriately combining those for the three dipole orientations at the location. It is evident that the large increase in LEE for the optimized \uled{}s is not simply due to a uniform increase in the LEE, but rather a nonuniform increase together with an enhancement in the dipole power in regions where the LEE is high. We can also observe sharp features in the LEE and emitted power. This suggests that simulations with few dipoles on a coarse grid may yield poor estimates of the total LEE.

\begin{figure}
    \centering
    \includegraphics{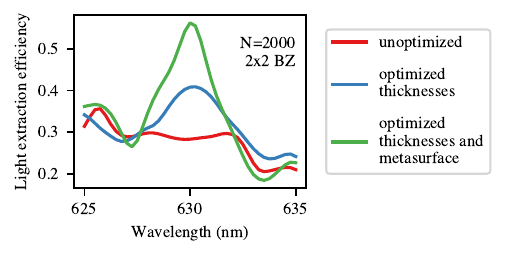}
    \caption{
    \label{fig:lee_vs_wavelength}
    Light extraction efficiency vs wavelength for the initial, unoptimized $\mu$LED; the $\mu$LED with optimized film thicknesses; and the $\mu$LED with optimized thicknesses and a metasurface.}
\end{figure}

Finally, \figref{fig:lee_vs_wavelength} shows the wavelength-dependent LEE for the three cases, and illustrates the strong enhancement of LEE at the 630 nm target wavelength. Away from the target, both optimized \uled{}s exhibit a significant drop in LEE, reaching values comparable to the unoptimized LED with less than 5 nm of wavelength shift. This suggests that a broadband objective will be required in order to obtain \uled{} designs which do not have strong wavelength dependence to LEE.

We emphasize that the \uled{}s discussed above are merely examples of designs that can be found through optimization, and we have seen many other designs with different metasurface densities and associated maps of LEE and emitted power. In practice, an objective should be crafted which is minimized for devices having the actual characteristics needed for \uled{} applications.

\section{Conclusion}
\label{section:conclusion}
We have shown that the FMM can be used to model \uled{} structures with accuracy comparable to high-resolution FDTD, while achieving a speedup greater than $10^7$ compared to CPU-based FDTD in situations where a large number of incoherent dipoles is considered. This result is enabled by the inherent characteristics of the FMM, vector FMM formulations that dramatically improve convergence compared to the basic FMM scheme, and BZ integration to model localized sources. These features are available in {\tt FMMAX}, a \emph{Jax}-based implementation of the FMM which we are open-sourcing alongside this work.

We used {\tt FMMAX} to optimize several \uled{} designs, and showed that a metasurface-enhanced \uled{} can increase LEE by 98\% over an unoptimized design, and by 38\% over a \uled{} with optimized film thicknesses but no metasurface. This result indicates that metasurfaces are a promising technology to improve \uled{} performance as required for applications such as AR displays. Next steps could include consideration of \uled{} structures that are more realistic, use of optimization objectives that better target the actual requirements of \uled{}s in AR applications (e.g. which maximize directional emission), the incorporation of manufacturability constraints in the inverse design scheme, and alternate initialization for \uled{} film thicknesses to obtain better local optima. It would also be interesting to study the new \emph{Jones direct} FMM formulation across a wider of structures.

Finally, our method could be applied to a range of problems where spatially incoherent emission is considered, to problems where localized sources interact with a periodic structure, or to problems where FMM is more typically employed, such as modeling of gratings or photonic crystals.

\pagebreak

\appendix
\section*{Appendix A: Vector field examples}
\label{appendix:vectors}

\begin{figure}
    \centering
    \includegraphics{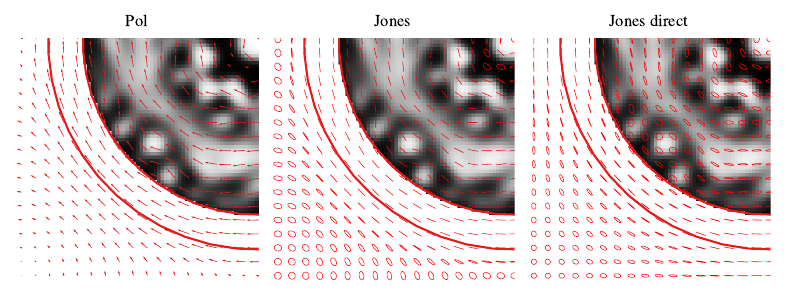}
    \caption{
    \label{fig:vector_fields}
    Automatically-generated vector fields for the metasurface layer in \figref{fig:metasurface}. \emph{Pol} yields a real-valued vector field tangent to material interfaces. \emph{Jones} yields a complex Jones vector field that is linear and tangent at interfaces, becoming circular away from interfaces. \emph{Jones direct} yields a field that is similar to \emph{Jones}---but critically, does not become circular between parallel interfaces, and is therefore smoother. The depicted vectors are sampled from an underlying, higher-resolution vector field.
    }
\end{figure}

Vector fields obtained for the metasurface layer in \figref{fig:metasurface} and the \emph{Pol}, \emph{Jones}, and \emph{Jones direct} methods are shown in \figref{fig:vector_fields}. The metasurface exhibits approximate four-fold rotational symmetry, and so we show one quadrant to make the details of vector fields more visible.

The \emph{Pol} field is obtained by solving Eq. \eqref{eqn:vector_optimization} starting with a real-valued $\mathbf{t}^\star$; the result is a real-valued vector field that at high-contrast interfaces is large and tangent to the interfaces. Zeros are located between parallel interfaces, e.g. at the top left, bottom left, and bottom right of the quadrant.

The \emph{Jones} field is obtained by converting the \emph{Pol} field into a complex Jones vector field as in \cite{LIU20122233}. Where the \emph{Pol} field has magnitude 1, the \emph{Jones} field corresponds to a linear polarization aligned with material interfaces. At zeros of the \emph{Pol} field, the \emph{Jones} field corresponds to a circular polarization.

The \emph{Jones direct} field is obtained by solving Eq. \eqref{eqn:vector_optimization} starting with a complex-valued $\mathbf{t}^\star$. To obtain the complex-valued $\mathbf{t}^\star$, we simply take the real-valued initial field and convert to a complex Jones vector field by the aforementioned procedure. The resulting Jones vector field again corresponds to a tangent linear polarization at high-contrast interfaces, and circular at certain locations (e.g. the bottom left of the quadrant), but which remains linear at some locations where the \emph{Jones} method produces a circular polarization (e.g. top left or bottom right of the quadrant). Thus, the \emph{Jones direct} field is (in some sense) smoother, and may be represented with fewer terms in a Fourier expansion---leading to improved convergence.

There are a few other differences between the \emph{Jones} and \emph{Jones direct} fields, e.g. locations where the \emph{Jones} field is fully linear and \emph{Jones direct} is merely elliptical. The field obtained by each of these methods can in principle be fine-tuned by adding hyperparameters to Eq. \eqref{eqn:vector_loss} which scale the smoothness and alignment terms. In future work, it would be interesting to study convergence in the context of this hyperparameter space and with a wider range of structures.

\section*{Appendix B: Comparison of FMM formulations}
\label{appendix:formulations}
{\tt FMMAX} implements several FMM formulations. These exhibit virtually identical convergence for the basic \uled{} structure of \figref{fig:structure}. However, for more complex \uled{}s that include metasurfaces this may not be expected. Further, in an inverse design setting one may encounter ``overfitting'' where an optimized design performs best when evaluated with simulation settings (i.e. number of Fourier orders, FMM formulation, and BZ integration scheme) used during the optimization procedure. This makes it challenging to identify a preferred formulation.

\begin{figure}
    \centering
    \includegraphics{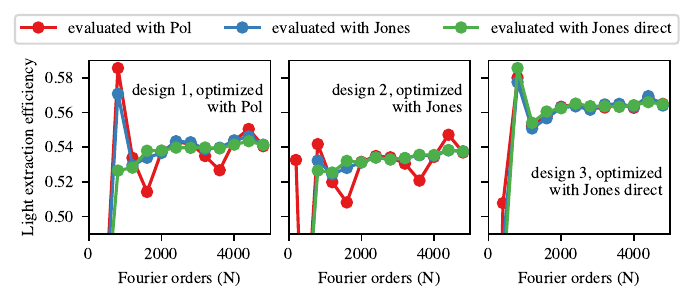}
    \caption{
    \label{fig:convergence_comparison}
    Light extraction efficiency as a function of Fourier orders $N$ for three separate metasurface-enhanced \uled{} designs. Each design was obtained by solving the inverse design problem of \secref{sec:optimization} using the \emph{Pol}, \emph{Jones}, and \emph{Jones direct} formulations during the course of optimization. We then evaluated each design with various $N$ and the three formulations. Overfitting to $N=800$ is seen in all cases, but the results converge as $N$ increases. \emph{Jones direct} exhibits the best convergence of the three formulations.
    }
\end{figure}

Thus, to select the formulation best suited for \uled{} design, we solved the inverse design problem from \secref{sec:optimization} with the \emph{Pol}, \emph{Jones}, and \emph{Jones direct} methods (with $N=800$), obtaining three separate designs. We then evaluated the total LEE of each design using the three formulations and Fourier orders up to $N=4800$. The results are shown in \figref{fig:convergence_comparison}.

The designs achieve similar performance and all show some degree of overfitting, where the simulation settings used for optimization yield a LEE higher than the converged result. However, increasing from $N=800$ to $N=2000$ (our validation setting, discussed in \secref{sec:convergence}) is sufficient to obtain converged results. Of the three formulations, \emph{Pol} converges least well, and exhibits some oscillation in the total LEE with an amplitude of $\sim$5\%. \emph{Jones} and \emph{Jones direct} are quite similar, with \emph{Jones direct} being slightly better on all three designs and avoiding the overfitting on the \emph{Pol}-optimized design. The results support the selection of \emph{Jones direct} and $N=2000$ for validation, although a more thorough study of convergence for a wider range of structures would be valuable.

\begin{backmatter}

% \bmsection{Funding}

% \bmsection{Acknowledgments}

\bmsection{Disclosures}
The authors declare no conflicts of interest.

% \bmsection{Data availability} Data underlying the results presented in this paper are available in [TODO].

\end{backmatter}

%%%%%%%%%%%%%%%%%%%%%%% References %%%%%%%%%%%%%%%%%%%%%%%%%
% Go after appendix and backmatter

\bibliography{references}

\begin{thebibliography}{10}
\newcommand{\enquote}[1]{``#1''}

\bibitem{Xiong2021}
J.~Xiong, E.-L. Hsiang, Z.~He, T.~Zhan, and S.-T. Wu, \enquote{Augmented
  reality and virtual reality displays: emerging technologies and future
  perspectives,} {\protect\JournalTitle{Light: Science {\&} Applications}}
  \textbf{10}, 216 (2021).

\bibitem{Park2021}
J.~Park, J.~H. Choi, K.~Kong, J.~H. Han, J.~H. Park, N.~Kim, E.~Lee, D.~Kim,
  J.~Kim, D.~Chung, S.~Jun, M.~Kim, E.~Yoon, J.~Shin, and S.~Hwang,
  \enquote{Electrically driven mid-submicrometre pixelation of {I}n{G}a{N}
  micro-light-emitting diode displays for augmented-reality glasses,}
  {\protect\JournalTitle{Nature Photonics}} \textbf{15}, 449--455 (2021).

\bibitem{Molesky2018}
S.~Molesky, Z.~Lin, A.~Y. Piggott, W.~Jin, J.~Vuckovi{\'{c}}, and A.~W.
  Rodriguez, \enquote{Inverse design in nanophotonics,}
  {\protect\JournalTitle{Nature Photonics}} \textbf{12}, 659--670 (2018).

\bibitem{meep_adjoint}
A.~M. Hammond, A.~Oskooi, M.~Chen, Z.~Lin, S.~G. Johnson, and S.~E. Ralph,
  \enquote{High-performance hybrid time/frequency-domain topology optimization
  for large-scale photonics inverse design,} {\protect\JournalTitle{Optics
  Express}} \textbf{30}, 4467--4491 (2022).

\bibitem{hammond_constraints}
A.~M. Hammond, A.~Oskooi, S.~G. Johnson, and S.~E. Ralph, \enquote{Photonic
  topology optimization with semiconductor-foundry design-rule constraints,}
  {\protect\JournalTitle{Optics Express}} \textbf{29}, 23916--23938 (2021).

\bibitem{MILONNI19761}
P.~Milonni, \enquote{Semiclassical and quantum-electrodynamical approaches in
  nonrelativistic radiation theory,} {\protect\JournalTitle{Physics Reports}}
  \textbf{25}, 1--81 (1976).

\bibitem{KIM1986396}
K.-J. Kim, \enquote{An analysis of self-amplified spontaneous emission,}
  {\protect\JournalTitle{Nuclear Instruments and Methods in Physics Research
  Section A: Accelerators, Spectrometers, Detectors and Associated Equipment}}
  \textbf{250}, 396--403 (1986).

\bibitem{POLIMERIDIS2015}
A.~G. Polimeridis, M.~T.~H. Reid, W.~Jin, S.~G. Johnson, J.~K. White, and A.~W.
  Rodriguez, \enquote{Fluctuating volume-current formulation of electromagnetic
  fluctuations in inhomogeneous media: Incandescence and luminescence in
  arbitrary geometries,} {\protect\JournalTitle{Phys. Rev. B}} \textbf{92},
  134202 (2015).

\bibitem{Carey2008}
V.~P. Carey, G.~Chen, C.~Grigoropoulos, M.~Kaviany, and A.~Majumdar, \enquote{A
  review of heat transfer physics,} {\protect\JournalTitle{Nanoscale and
  Microscale Thermophysical Engineering}} \textbf{12}, 1--60 (2008).

\bibitem{Basu2009}
S.~Basu, Z.~M. Zhang, and C.~J. Fu, \enquote{Review of near-field thermal
  radiation and its application to energy conversion,}
  {\protect\JournalTitle{International Journal of Energy Research}}
  \textbf{33}, 1203--1232 (2009).

\bibitem{cylindrical_uled_fdtd}
H.-Y. Ryu, J.~Pyo, and H.~Y. Ryu, \enquote{Light extraction efficiency of
  {G}a{N}-based micro-scale light-emitting diodes investigated using
  finite-difference time-domain simulation,} {\protect\JournalTitle{IEEE
  Photonics Journal}} \textbf{12}, 1--10 (2020).

\bibitem{Chung:22}
H.~Chung, \enquote{Computational upper-limit of directional light emission in
  nano-led via inverse design,} {\protect\JournalTitle{Opt. Express}}
  \textbf{30}, 9008--9020 (2022).

\bibitem{multisource}
H.-C. Lin, Z.~Wang, and C.~W. Hsu, \enquote{Fast multi-source nanophotonic
  simulations using augmented partial factorization,}
  {\protect\JournalTitle{Nature Computational Science}} \textbf{2}, 815--822
  (2022).

\bibitem{li2023fast}
S.~Li, H.-C. Lin, and C.~W. Hsu, \enquote{Fast multi-channel inverse design
  through augmented partial factorization,}  (2023).

\bibitem{Yao2022}
W.~Yao, F.~Verdugo, R.~E. Christiansen, and S.~G. Johnson, \enquote{Trace
  formulation for photonic inverse design with incoherent sources,}
  {\protect\JournalTitle{Structural and Multidisciplinary Optimization}}
  \textbf{65}, 336 (2022).

\bibitem{Li:93}
L.~Li and C.~W. Haggans, \enquote{Convergence of the coupled-wave method for
  metallic lamellar diffraction gratings,} {\protect\JournalTitle{J. Opt. Soc.
  Am. A}} \textbf{10}, 1184--1189 (1993).

\bibitem{BZ_integration_1}
E.~López-Fraguas, F.~Binkowski, S.~Burger, S.~Hagedorn, B.~García-Cámara,
  R.~Vergaz, C.~Becker, and P.~Manley, \enquote{Tripling the light extraction
  efficiency of a deep ultraviolet led using a nanostructured p-contact,}
  {\protect\JournalTitle{Scientific Reports}} \textbf{12}, 11480 (2022).

\bibitem{BZ_integration_2}
L.~Zschiedrich, H.~J. Greiner, S.~Burger, and F.~Schmidt, \enquote{Numerical
  analysis of nanostructures for enhanced light extraction from {OLED}s,}
  (2013). Presented at Light-Emitting Diodes: Materials, Devices, and
  Applications for Solid State Lighting XVII.

\bibitem{BZ_integration_3}
L.~Zschiedrich, L.~Blome, and H.~J. Greiner, \enquote{Simulation of advanced
  {OLED} light extraction structures with novel {FEM} methods,}  (2014).
  Presented at Organic Photonics VI.

\bibitem{jax2018github}
J.~Bradbury, R.~Frostig, P.~Hawkins, M.~J. Johnson, C.~Leary, D.~Maclaurin,
  G.~Necula, A.~Paszke, J.~Vander{P}las, S.~Wanderman-{M}ilne, and Q.~Zhang,
  \enquote{{JAX}: Composable transformations of {P}ython+{N}um{P}y programs,}
  \url{http://github.com/google/jax} (2018).

\bibitem{KIM2023108552}
C.~Kim and B.~Lee, \enquote{{TORCWA}: {GPU}-accelerated {F}ourier modal method
  and gradient-based optimization for metasurface design,}
  {\protect\JournalTitle{Computer Physics Communications}} \textbf{282}, 108552
  (2023).

\bibitem{Colburn:21}
S.~Colburn and A.~Majumdar, \enquote{Inverse design and flexible
  parameterization of meta-optics using algorithmic differentiation,}
  {\protect\JournalTitle{Communications Physics}} \textbf{4}, 65 (2021).

\bibitem{Jin2020}
W.~Jin, W.~Li, M.~Orenstein, and S.~Fan, \enquote{Inverse design of lightweight
  broadband reflector for relativistic lightsail propulsion,}
  {\protect\JournalTitle{ACS Photonics}} \textbf{7}, 2350--2355 (2020).

\bibitem{nannos}
B.~Vial, \enquote{{nannos}: {F}ourier modal method for multilayer
  metamaterials,} \url{http://github.com/benvial/nannos} (2021).

\bibitem{LIU20122233}
V.~Liu and S.~Fan, \enquote{S4 : A free electromagnetic solver for layered
  periodic structures,} {\protect\JournalTitle{Computer Physics
  Communications}} \textbf{183}, 2233--2244 (2012).

\bibitem{Whittaker:09}
D.~M. Whittaker and I.~S. Culshaw, \enquote{Scattering-matrix treatment of
  patterned multilayer photonic structures,} {\protect\JournalTitle{Phys. Rev.
  B}} \textbf{60}, 2610 (1999).

\bibitem{Li:96}
L.~Li, \enquote{Use of {F}ourier series in the analysis of discontinuous
  periodic structures,} {\protect\JournalTitle{J. Opt. Soc. Am. A}}
  \textbf{13}, 1870--1876 (1996).

\bibitem{Gotz:08}
P.~G\"{o}tz, T.~Schuster, K.~Frenner, S.~Rafler, and W.~Osten, \enquote{Normal
  vector method for the {RCWA} with automated vector field generation,}
  {\protect\JournalTitle{Opt. Express}} \textbf{16}, 17295--17301 (2008).

\bibitem{Antos:09}
R.~Antos, \enquote{Fourier factorization with complex polarization bases in
  modeling optics of discontinuous bi-periodic structures,}
  {\protect\JournalTitle{Opt. Express}} \textbf{17}, 7269--7274 (2009).

\bibitem{Lalanne:00}
P.~Lalanne and E.~Silberstein, \enquote{Fourier-modal methods applied to
  waveguide computational problems,} {\protect\JournalTitle{Opt. Lett.}}
  \textbf{25}, 1092--1094 (2000).

\bibitem{Silberstein:01}
E.~Silberstein, P.~Lalanne, J.-P. Hugonin, and Q.~Cao, \enquote{Use of grating
  theories in integrated optics,} {\protect\JournalTitle{J. Opt. Soc. Am. A}}
  \textbf{18}, 2865--2875 (2001).

\bibitem{band_structure_engineering}
A.~Moritz and A.~Hangleiter, \enquote{{Optical gain in ordered GaInP/AlGaInP
  quantum wells},} {\protect\JournalTitle{Applied Physics Letters}}
  \textbf{66}, 3340--3342 (1995).

\bibitem{sigmund_overview}
O.~Sigmund and K.~Maute, \enquote{Topology optimization approaches,}
  {\protect\JournalTitle{Structural and Multidisciplinary Optimization}}
  \textbf{48}, 1031--1055 (2013).

\bibitem{CHRISTIANSEN201923}
R.~E. Christiansen, J.~Vester-Petersen, S.~P. Madsen, and O.~Sigmund,
  \enquote{A non-linear material interpolation for design of metallic
  nano-particles using topology optimization,} {\protect\JournalTitle{Computer
  Methods in Applied Mechanics and Engineering}} \textbf{343}, 23--39 (2019).

\bibitem{svanberg2013density}
K.~Svanberg and H.~Sv{\"a}rd, \enquote{Density filters for topology
  optimization based on the pythagorean means,}
  {\protect\JournalTitle{Structural and Multidisciplinary Optimization}}
  \textbf{48}, 859--875 (2013).

\bibitem{lbfgsb}
R.~H. Byrd, P.~Lu, J.~Nocedal, and C.~Zhu, \enquote{A limited memory algorithm
  for bound constrained optimization,} {\protect\JournalTitle{SIAM Journal on
  Scientific Computing}} \textbf{16}, 1190--1208 (1995).

\bibitem{meep}
A.~F. Oskooi, D.~Roundy, M.~Ibanescu, P.~Bermel, J.~D. Joannopoulos, and S.~G.
  Johnson, \enquote{{MEEP}: A flexible free-software package for
  electromagnetic simulations by the {FDTD} method,}
  {\protect\JournalTitle{Computer Physics Communications}} \textbf{181},
  687--702 (2010).

\bibitem{nlopt}
S.~G. Johnson, \enquote{The {NLopt} nonlinear-optimization package,}
  http://github.com/stevengj/nlopt.

\bibitem{taflove}
A.~Taflove, S.~C. Hagness, and M.~Piket-May, \enquote{Computational
  electromagnetics: the finite-difference time-domain method,}
  {\protect\JournalTitle{The Electrical Engineering Handbook}} \textbf{3},
  629--670 (2005).

\bibitem{deinega2007subpixel}
A.~Deinega and I.~Valuev, \enquote{Subpixel smoothing for conductive and
  dispersive media in the finite-difference time-domain method,}
  {\protect\JournalTitle{Optics letters}} \textbf{32}, 3429--3431 (2007).

\bibitem{oskooi_factorized}
A.~Oskooi, C.~Hogan, A.~M. Hammond, M.~Reid, and S.~G. Johnson,
  \enquote{Factorized machine learning for performance modeling of massively
  parallel heterogeneous physical simulations,} {\protect\JournalTitle{arXiv
  preprint arXiv:2003.04287}}  (2020).

\end{thebibliography}

\end{document}